\documentclass[longbibliography,reprint,aps,prb,citeautoscript,superscriptaddress,10pt]{revtex4-1}

\usepackage[utf8]{inputenc}
\usepackage{graphicx}
\usepackage{amsmath}
\usepackage{hyperref}

\usepackage[T1]{fontenc}
\usepackage{lmodern}

\begin{document}

\title{Disorder raises the critical temperature of a cuprate superconductor}

\author{Maxime Leroux}
\altaffiliation{Present address: MPA-CMMS, Los Alamos National Laboratory, Los Alamos, NM 87545, USA}
\affiliation{Materials Science Division, Argonne National Laboratory, 9700 Cass Avenue, Lemont, IL 60439, USA}

\author{Vivek Mishra}
\altaffiliation{Present address: Computational Sciences and Engineering Division and Center for Nanophase Materials Sciences, Oak Ridge National Laboratory, Oak Ridge, TN 37831}
\affiliation{Materials Science Division, Argonne National Laboratory, 9700 Cass Avenue, Lemont, IL 60439, USA}

\author{Jacob P.C. Ruff}
\affiliation{CHESS, Cornell University, Ithaca, NY 14853, USA}

\author{Helmut Claus}
\affiliation{Materials Science Division, Argonne National Laboratory, 9700 Cass Avenue, Lemont, IL 60439, USA}

\author{Matthew P. Smylie}
\altaffiliation{Department of Physics and Astronomy, Hofstra University, Hempstead, NY 11549}
\affiliation{Materials Science Division, Argonne National Laboratory, 9700 Cass Avenue, Lemont, IL 60439, USA}

\author{Christine Opagiste}
\affiliation{Institut NÉEL, CNRS, Univ. Grenoble Alpes, 25 Avenue des Martyrs, 38042 Grenoble, France}

\author{Pierre Rodière}
\affiliation{Institut NÉEL, CNRS, Univ. Grenoble Alpes, 25 Avenue des Martyrs, 38042 Grenoble, France}

\author{Asghar Kayani}
\affiliation{Department of Physics, Western Michigan University, Kalamazoo, MI 49008, USA}

\author{G. D. Gu}
\affiliation{Condensed Matter Physics and Materials Science Department, Brookhaven National Laboratory, Upton, NY 11973-5000, USA}

\author{John M. Tranquada}
\affiliation{Condensed Matter Physics and Materials Science Department, Brookhaven National Laboratory, Upton, NY 11973-5000, USA}

\author{Wai-Kwong Kwok}
\affiliation{Materials Science Division, Argonne National Laboratory, 9700 Cass Avenue, Lemont, IL 60439, USA}

\author{Zahirul Islam}
\affiliation{Advanced Photon Source, Argonne National Laboratory, 9700 Cass Avenue, Lemont, IL 60439, USA}

\author{Ulrich Welp}
\affiliation{Materials Science Division, Argonne National Laboratory, 9700 Cass Avenue, Lemont, IL 60439, USA}
\email{welp@anl.gov }



\newcommand{\Tc}{$T_\mathrm{c}$}
\newcommand{\TCDW}{$T_\mathrm{CDW}$}

\newcommand{\LBCOx}{La$_{2-x}$Ba$_x$CuO$_4$}
\newcommand{\LNSCOx}{La$_{1.6-x}$Nd$_{0.4}$Sr$_x$CuO$_4$}
\newcommand{\LBCO}{La$_{1.875}$Ba$_{0.125}$CuO$_4$}
\newcommand{\LBSCOx}{La$_{1.875}$Ba$_{0.125-x}$Sr$_x$CuO$_4$}
\newcommand{\YBCOd}{YBa$_2$Cu$_3$O$_{7-\delta}$}
\newcommand{\BSCCOx}{Bi$_2$Sr$_2$CaCu$2$O$_{8+x}$}
\newcommand{\HBCOd}{HgBa$_2$CuO$_{4+\delta}$}
\newcommand{\LIS}{Lu$_5$Ir$_4$Si$_{10}$}






\begin{abstract}
%
With the discovery of charge density waves (CDW) in most members of the cuprate high temperature superconductors, the interplay between superconductivity and CDW has become a key point in the debate on the origin of high temperature superconductivity.
Some experiments in cuprates point toward a CDW state competing with superconductivity, but others raise the possibility of a CDW-superconductivity intertwined order, or more elusive pair-density wave (PDW).
Here we have used proton irradiation to induce disorder in crystals of \LBCO~and observed a striking 50\% \emph{increase} of \Tc\ accompanied by a suppression of the CDW. This is in sharp contrast with the behavior expected of a d-wave superconductor for which both magnetic and non-magnetic defects should suppress \Tc. Our results thus make an unambiguous case for the \emph{strong} detrimental effect of the CDW on bulk superconductivity in \LBCO. Using tunnel diode oscillator (TDO) measurements, we find indications for potential dynamic layer decoupling in a PDW phase. Our results establish irradiation-induced disorder as a particularly relevant tuning parameter for the many families of superconductors with coexisting density waves, which we demonstrate on superconductors such as the dichalcogenides and \LIS.
\end{abstract}

\maketitle

\textbf{Significance statement\\
The origin of high-temperature superconductivity remains unknown. Over the past two decades, spatial oscillations of the electronic density known as Charge Density Waves (CDW) have been found to coexist with high-temperature superconductivity in most prominent cuprate superconductors. The debate on whether CDW help or hinder high-temperature superconductivity in cuprates is still ongoing. In principle, disorder at the atomic scale should strongly suppress both high-temperature superconductivity and CDW. In this article however, we find that disorder created by irradiation increases the superconducting critical temperature by 50\% while suppressing the CDW order, showing that CDW \emph{strongly} hinder bulk superconductivity. We provide indications this increase occurs because the CDW could be part of a pair-density wave frustrating the superconducting coupling between atomic planes.}

\begin{figure}[t!]
\includegraphics[width=\linewidth]{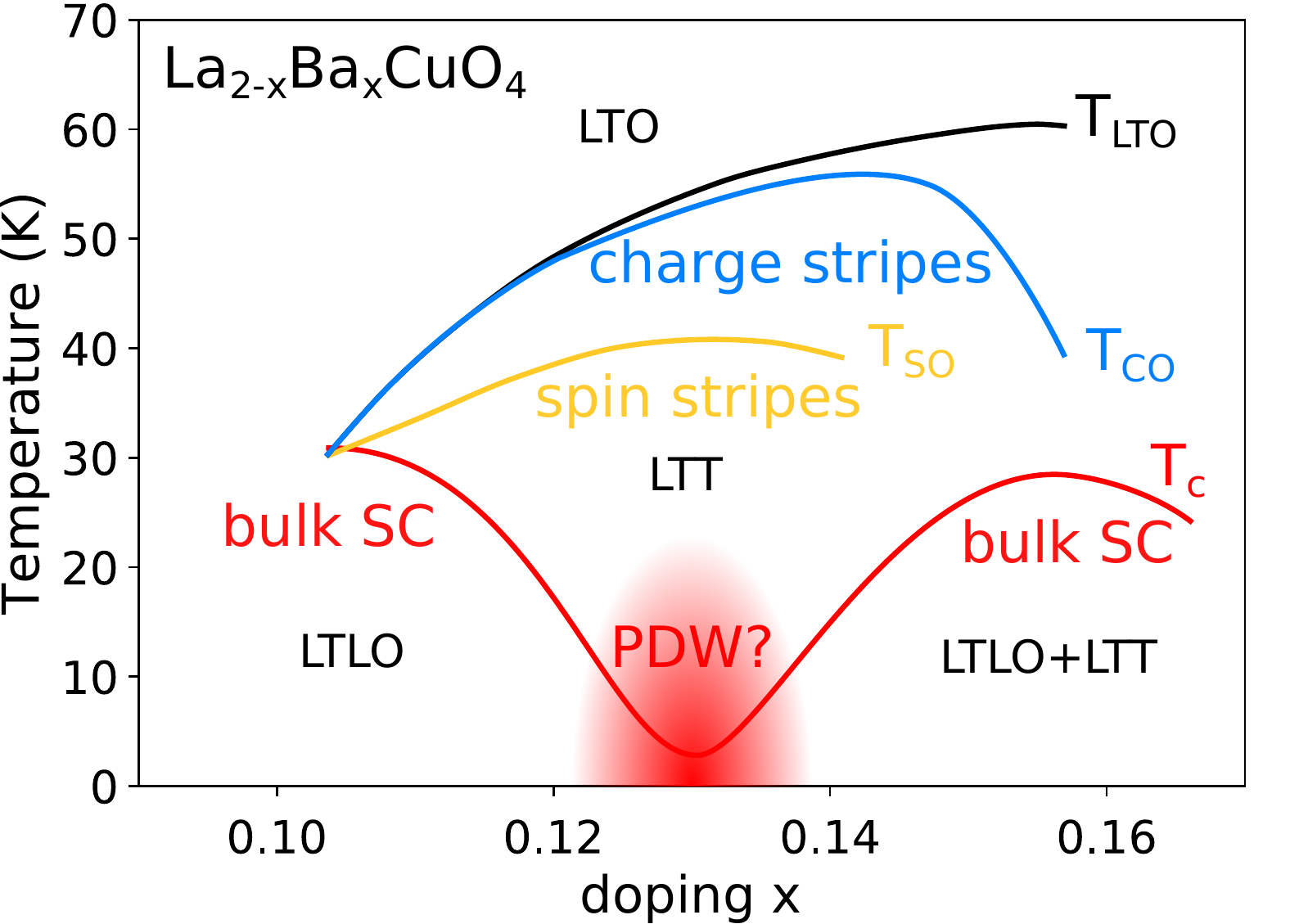}
\caption{\textbf{Schematic phase diagram of \LBCOx.}
A marked suppression of the superconducting critical temperature (\Tc) occurs around 1/8 doping.
At this doping level charge and spin stripe order is the most pronounced. It has been proposed that the stripe order is accompanied by a pair-density wave (PDW) state. 
Structural phases: low-temperature orthorhombic (LTO), low-temperature
tetragonal (LTT), low-temperature less-orthorhombic (LTLO). Data adapted from Ref.~\cite{Hucker2011EXSLBCOstripe} }
\label{fig:LBCO_schemepd}
\end{figure}

Charge density waves (CDW) are real-space periodic oscillations of the crystal electronic density accompanied by a lattice distortion. In their simplest form, CDW can be described as a Bardeen-Cooper-Schrieffer condensate of electron-hole pairs that breaks the translational symmetry\cite{gruner2000density}.
CDW in cuprate superconductors were first observed in lanthanum based compounds\cite{Tranquada1995stripenature,Tranquada1996LNSCOprl,Fujita2002LBSCOcdwsdw}, such as \LBCOx~(LBCO) around 1/8 hole-doping\cite{Tranquada1995stripenature,Fujita2002LBSCOcdwsdw,Kim2008LBCOstructCO,Hucker2011EXSLBCOstripe,JieTranqQiangLi2012stripetransport,MiaoLBCOCDWnolockin,RajasekaranCavalleriScience2018,GuguchiaPRLLBCOimpurity,bednorz1986possible, Fujita2004pureLBCOstripes}.
In these La-based cuprates, spin correlations were also found to synchronize with the CDW modulation to form static ``stripes''\cite{Tranquada1995stripenature} of interlocked CDW and spin density waves (SDW) at a certain temperature below the ordering temperature of the CDW (see Fig.~\ref{fig:LBCO_schemepd}).

More recently, CDW were shown to be a nearly-universal characteristic in cuprates: a CDW was observed in hole-doped \YBCOd~(YBCO)\cite{WuJulien2011CDWNMR,Wu2013,Wu2015,LeTacon2013IXSYBCONatPhys,Chang2012YBCOcompetingCDW,Gerber3DCDWYBCOhighB,JangYBCOCDWhighfield}
as well as several others hole-doped cuprates\cite{NetoGuTacon2014Science,Tabis2014NatComHgBCO_cdw}, and it was also found in electron-doped Nd$_{2-x}$Ce$_x$CuO$_4$\cite{daSilvaNetoCDWelecdopedCuprate}.
These CDW seemed at first notably different from the stripes in La-based cuprates, as they were not accompanied by a SDW. But recent studies suggest that the three-dimensional CDW induced by magnetic field in YBCO might have connections with LBCO in terms of a pair-density wave (PDW) state\cite{KacmarcikKlein2018,YuPNAS2016YBCOunderdoped}.

The role of these density waves in the bigger picture of high temperature superconductivity is subject to intense and active research\cite{FradkinTranquada2015RoMP}.
To first order, CDW and superconductivity compete for the same electronic density of states near the Fermi level as was observed in several s-wave CDW superconductors such as 2H-TaS/Se$_2$\cite{Mutka1983PRB_CDW_SC_irrad} and \LIS\cite{Shelton1986LIS_P_rho_Cp,SinghPRB2005,Leroux2013LIS_SCprop}.
Indeed, in cuprate superconductors, the current experimental evidence seems to point toward a scenario, where the CDW state is in competition with bulk superconductivity. The superconducting \Tc\ exhibits a plateau\cite{Ramshaw317,CAVA1988YBCO60Kplateau}
or a deep minimum\cite{Hucker2011EXSLBCOstripe,MoodenbaughPRBLBCO2Tcdomes}
in the range of doping where the CDW occurs. X-ray studies in YBCO also found that the CDW phase is suppressed when entering the superconducting phase\cite{LeTacon2013IXSYBCONatPhys}. Conversely, the CDW was found to be enhanced when superconductivity is suppressed by a magnetic field\cite{Chang2012YBCOcompetingCDW,Kim2008PRBLBCO_stripe_magneticfield,Gerber3DCDWYBCOhighB,JangYBCOCDWhighfield},
and the superconducting critical temperature can be restored when the CDW is suppressed under pressure\cite{IdoJLTP1996pressureeffectonLBCO_LNBCO}. 
Despite this range of results in favor of competition, a definitive consensus has not yet been reached as the tuning parameters mentioned above may have non trivial beneficial or detrimental effects on both superconductivity and CDW, considering the complex mix of charge, spin and electronic orders involved.

Moreover, there is evidence that the fate of superconductivity and CDW are intertwined in La-based cuprates with stripes. Angle resolved photo-emission spectroscopy (ARPES) data is consistent with a phase-incoherent d-wave superconductor whose Cooper pairs form spin-charge ordered structures instead of becoming superconducting\cite{Valla1914}. Extremely small in-plane resistance was also found in the stripe phase above the superconducting \Tc\cite{QLiPRL2007,BergPRL2007LBCOTcT3DTKT}. Further evidence for ``in-plane'' superconductivity in the normal-state striped phase of LBCO (above the bulk superconducting \Tc) was found in the non-linear optical response, with clear signatures of superconducting tunneling above \Tc\ and up to the CDW transition temperature\cite{RajasekaranCavalleriScience2018}. A recovery of the interlayer Josephson coupling, marked by a rapid and transient blue-shift of the Josephson plasmon resonance, was also observed when the stripe order is melted by near-infrared femtosecond laser pulses\cite{KhannaCavalleri2016}.
It was also shown that the spin stripe ordering temperature is suppressed in the same manner as the superconducting transition temperature when doping with Zn\cite{GuguchiaPRLLBCOimpurity}, which strongly suggests that the existence of stripes requires intertwining with superconducting pairing correlations.
Non-magnetic Zn impurities are indeed known to suppress \Tc\ in cuprates\cite{FukuzumiPRLTccupratereducebyZn,AlloulHirschfeldRMPdefects}.

A leading hypothesis to explain these anomalies is the so-called PDW state: a condensate of Cooper pairs (electron-electron pairs) with finite center of mass momentum, giving rise to real space modulations with no uniform component\cite{BergPRL2007LBCOTcT3DTKT,BergTranquada2009NJoP,FradkinTranquada2015RoMP,Tranquada2008PDW_PRB}. In a recent STM experiment, detection of PDW order in vortex cores has further supported the idea of PDW's relevance for cuprates\cite{EdkinsPDWSTM}.
Because stripes are oriented perpendicular to one another in adjacent planes in cuprates, a PDW would have no first-order Josephson coupling between adjacent planes, in agreement with the non-linear optical response\cite{KhannaCavalleri2016,RajasekaranCavalleriScience2018}. The PDW's inherent frustration of the Josephson coupling would also yield zero in-plane resistance but no bulk Meissner effect and zero critical current\cite{BergTranquada2009NJoP}, in agreement with transport and magnetization studies\cite{QLiPRL2007,BergPRL2007LBCOTcT3DTKT}.
As CDW and PDW are modulated in real-space they should have a strong sensitivity to disorder\cite{BergTranquada2009NJoP}.

In this article, we prove that CDW and bulk superconductivity are \emph{strongly} competing in LBCO close to 1/8 doping. In a first for cuprates, we observe a striking \emph{increase} of \Tc\ and suppression of the CDW after increasing the disorder via irradiation, in sharp contrast with the expected behavior of a d-wave superconductor.
Beyond the competing CDW scenario, our results would also have a natural explanation in the PDW scenario, in terms of a restored Josephson coupling. This would imply that the PDW is impeding the bulk d-wave superconductivity in LBCO.
More generally, our results establish irradiation-induced disorder as a particularly relevant tuning parameter to study the many families of superconductors that have coexisting density waves and other spatially modulated orders.

\begin{figure}[t!]
\includegraphics[width=\linewidth]{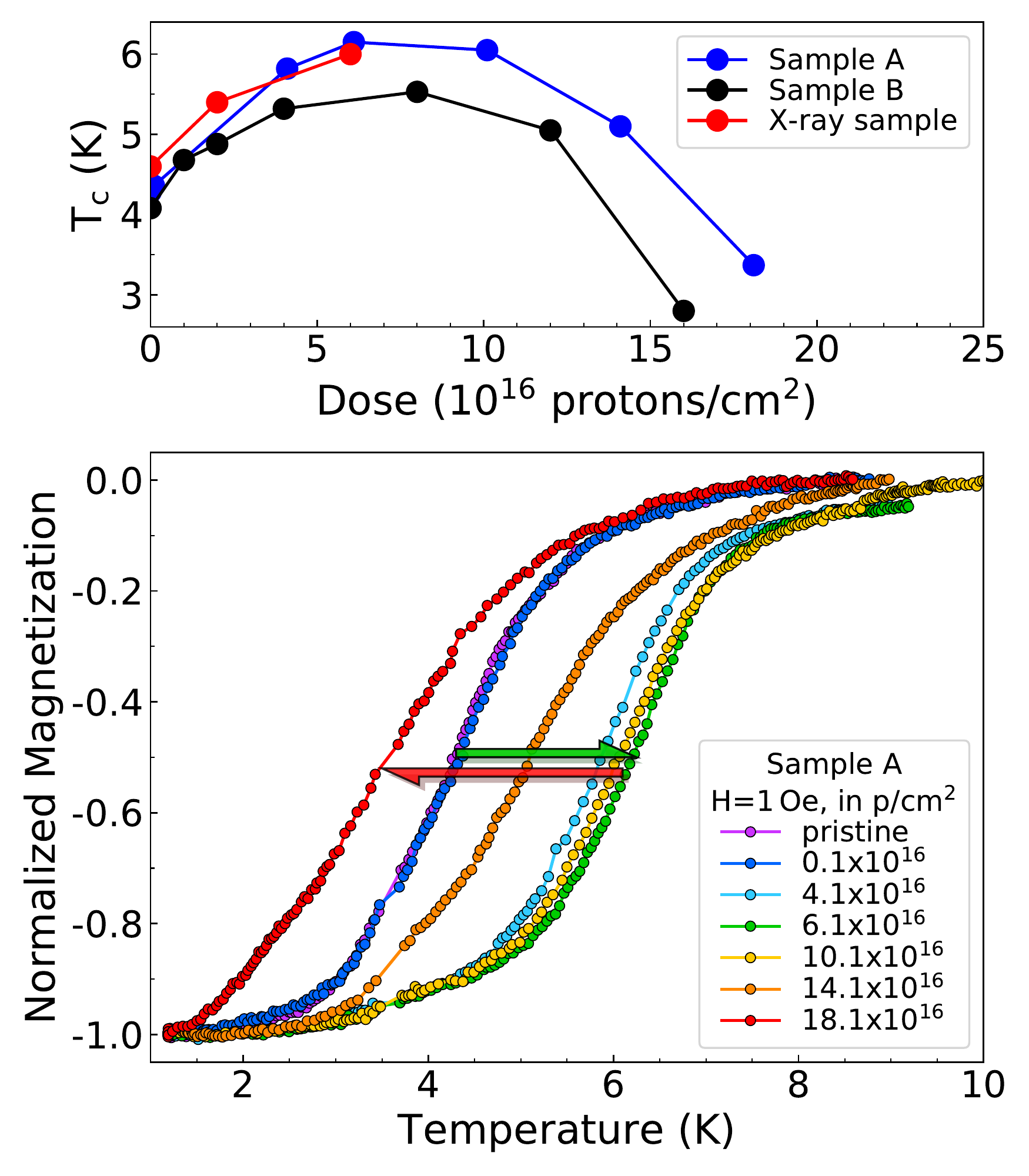}
\caption{\textbf{Irradiation enhances superconductivity in \LBCO.}
\textbf{(top)} Evolution of the superconducting transition temperature \Tc\ (defined as a 50\% reduction in magnetization to a complete Meissner effect) of several samples of \LBCO, as a function of irradiation dose.\textbf{(bottom)} Normalized magnetization of \LBCO~showing the temperature evolution of the Meissner effect, as a function of 5-MeV proton-irradiation dose. A clear increase of \Tc\ is observed, in contrast with the expected behavior of a normal d-wave superconductor. The transition temperature, which starts out at 4.28 K, peaks at 6.15 K after irradiation to $6.1\times10^{16}$\,p/cm$^2$, and then falls back to 3.37 K after the last irradiation. The \Tc\ curves stay sharp after the first four irradiations, indicative of uniform superconducting properties and uniform irradiation damage.
}
\label{fig:LBCO_curve}
\end{figure}

\section{RESULTS}

In Fig.~\ref{fig:LBCO_curve} we report magnetization measurements showing the expulsion of the magnetic field (Meissner effect) in a $75 \,\mathrm{\mu m}$-thick 1/8-doped LBCO single crystal which was repeatedly irradiated with 5\,MeV protons. The bulk \Tc, defined as the mid-point of the transition to a complete Meissner effect, is enhanced by up to 50\% as the cumulative irradiation dose increases. This behavior was consistently observed across several samples of LBCO, as reported in the upper panel of Fig.~\ref{fig:LBCO_curve}. Notice also how up to $10\times10^{16}$ p/cm$^2$ the magnetization curves stay sharp and parallel. The curves simply shift uniformly to higher and higher temperature after irradiation, indicating that the superconducting properties stay uniform. For the two highest irradiation doses ($14.1$ and $18.1\times10^{16}$\,p/cm$^2$) the curves finally start to broaden and shift back to lower temperature, such high doses are comparable to those known to degrade cuprate superconductors properties\cite{Jia2013protonirrad}.

As mentioned in the introduction and Fig.~\ref{fig:LBCO_schemepd}, LBCO has charge and spin stripe orders and undergoes a series of separate transition as a function of temperature\cite{BergPRL2007LBCOTcT3DTKT}: the resistivity of the ab-plane
drops below $T_{SDW}=42\,K$ and becomes immeasurably small ($<10^{-4}\,$m$\Omega$.cm) at $T_{KT}\approx16$\,K.
In between, the dependence is qualitatively of the Berezinskii-Kosterlitz-Thouless form suggesting superconducting fluctuations restricted to individual copper-oxide planes. The resistivity of the c-axis also vanishes but only below $T_{3D}\approx10$\,K, so that, within experimental errors, the anisotropy of the resistivity is infinite between $T_{3D}$ and $T_{KT}$. Finally, the Meissner effect occurs below 4\,K which has been considered as the true \Tc\cite{QLiPRL2007,BergPRL2007LBCOTcT3DTKT}. We performed a series of complimentary measurements which agree with the magnetization measurements and reveal a richer picture.
\begin{figure}[t!]
\includegraphics[width=\linewidth]{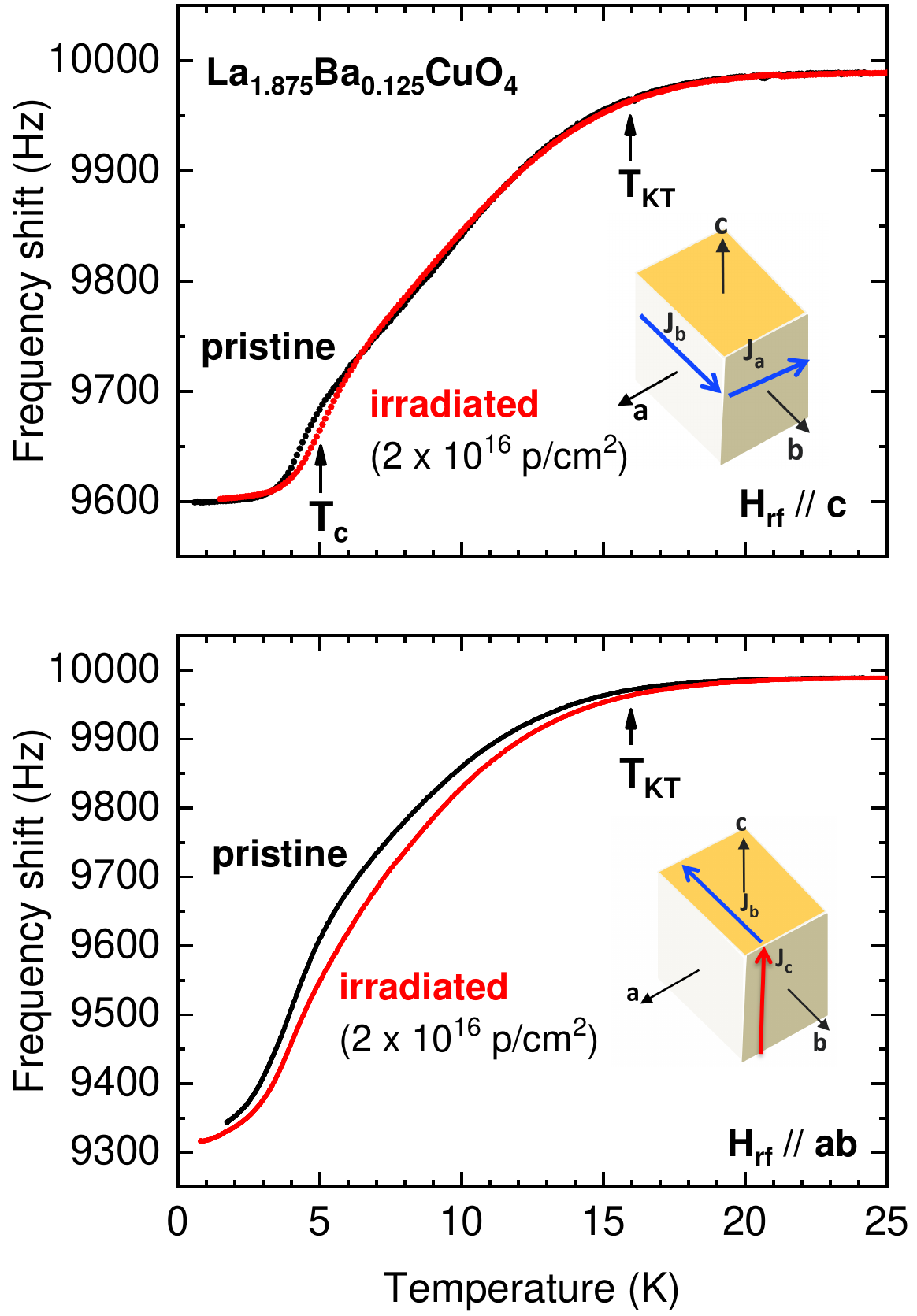}
\caption{\textbf{Tunnel diode oscillator measurements show enhanced superconductivity after irradiation}
The frequency shift of the oscillator is proportional to the superconducting magnetic penetration depth ($\lambda$). \textbf{(top)} When parallel to the c-axis (see inset), the small ($\mu$T) oscillating magnetic field probes mostly the in-plane $\lambda_{ab}$. Below 5\,K, the small shift toward {higher} temperature after irradiation is in agreement with the change in Meissner effect reported in Fig.\ref{fig:LBCO_curve}. \textbf{(bottom)} When the field is aligned in-plane, it probes both $\lambda_{ab}$ and $\lambda_{c}$. In this case, we observe a global shift toward higher temperature, which is suggestive of enhanced c-axis Josephson coupling. The Kosterlitz-Thouless (KT) temperature reported in \cite{BergPRL2007LBCOTcT3DTKT} is also indicated by an arrow.
}
\label{fig:LBCO_TDO}
\end{figure}

In Fig.~\ref{fig:LBCO_TDO} we show the frequency shift in tunnel diode oscillator measurements, which is proportional to the length over which the external magnetic field is screened\cite{Smylie2016PRBtopinsul,Smylie2016PRBprotweakpairbreak}. 
In the normal state, this is the skin depth ($\delta$), and in the superconducting state it is the magnetic penetration depth ($\lambda$).
The small ($\approx 5\,\mu$T) magnetic field oscillating near 14\,MHz mostly probes the in-plane $\lambda_{ab}$ when applied along the c-axis (Fig.~\ref{fig:LBCO_TDO} upper panel).
The frequency starts to decrease below 18\,K, where the in-plane resistivity was reported to tend toward zero\cite{QLiPRL2007}, and in agreement with previous studies\cite{BergPRL2007LBCOTcT3DTKT} which provide evidence for the onset of in-plane superconductivity below $T_{KT}\approx16\,$K.
Then a faster decrease occurs below 5\,K where the Meissner effect was observed in magnetization measurements.
Strikingly, only the part below 5\,K shifts toward \emph{higher} temperature after irradiation and the midpoint of this transition shifts from 4.5 to 5.0\,K for $2\times10^{16}$ p/cm$^2$, in excellent agreement with the magnetization measurements of Fig.\ref{fig:LBCO_curve}.
When the field is applied in-plane, both $\lambda_{ab}$ and $\lambda_{c}$ are probed.
Again, the frequency starts to decrease below 18\,K and decreases faster below 5\,K, but there is a global shift toward higher temperature after irradiation, which is suggestive of enhanced c-axis superconducting Josephson coupling, as we will discuss later.

\begin{figure}
\centering
\includegraphics[width=\linewidth]{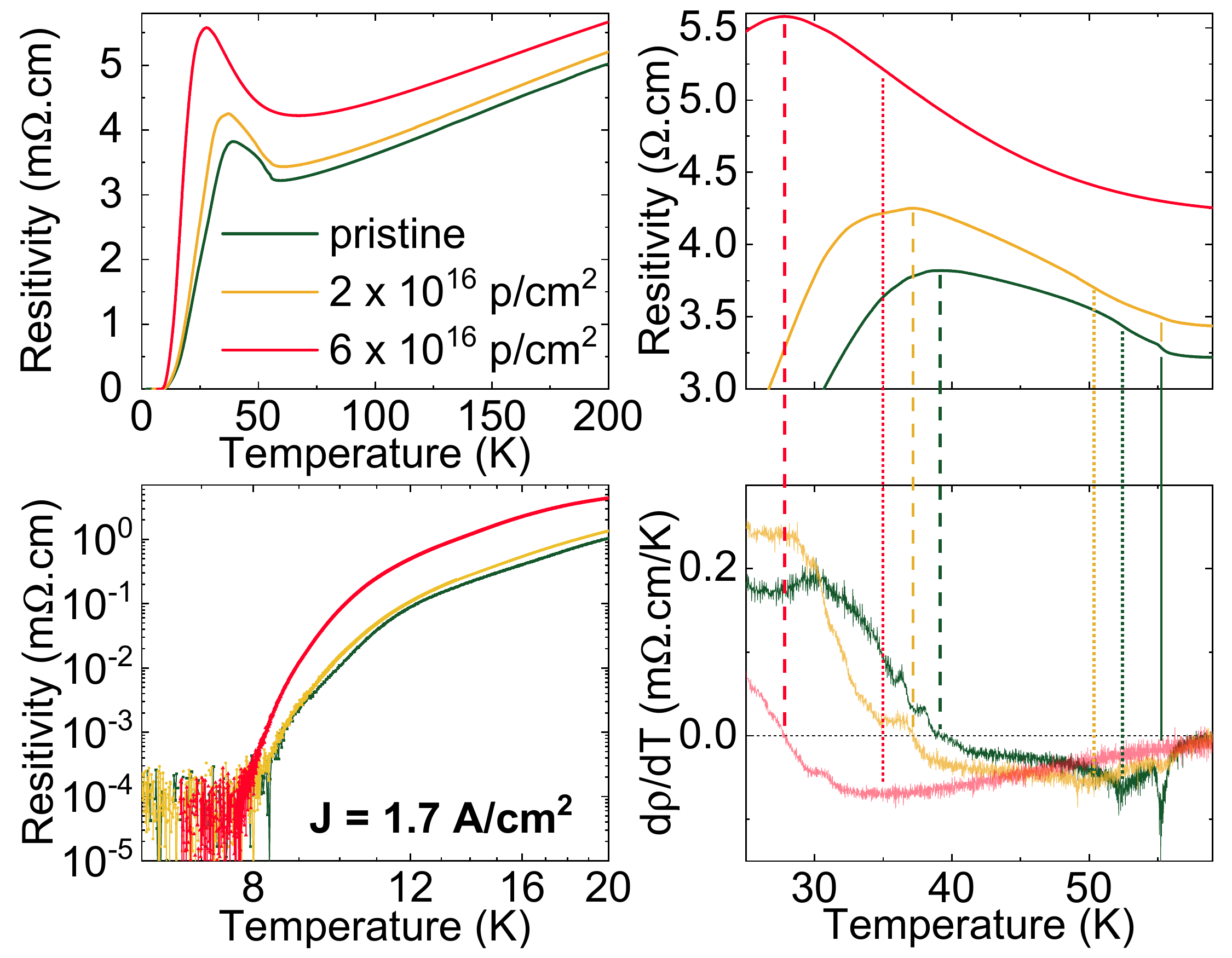}
\caption{\textbf{Resistivity suggests suppressed charge order after irradiation}
\textbf{(top-left)} Temperature dependence of the in-plane resistivity of LBCO (with a small c-axis component, see text) as a function of irradiation dose. Increased disorder yields the expected global upward shift. \textbf{(bottom-left)} Log-log plot of the resistivity near the superconducting transition. As previously reported\cite{BergPRL2007LBCOTcT3DTKT}, the resistivity drops rapidly a few kelvins above the Meissner effect at 4\,K. The resistivity gets steeper after irradiation, suggesting a possible crossing below the sensitivity of our instruments (10$^{-7}\,\Omega$.cm). \textbf{(right)} Resistivity and its derivative as a function of temperature around the structural and CDW transitions. A solid line indicates the structural transition at 55.3\,K, dotted lines indicate the inflection point at a temperature immediately below which seems to match the CDW transition, dashed lines indicate the maximum resistivity.
}
\label{fig:LBCO_res}
\end{figure}
We also measured the temperature dependence of the in-plane resistivity of another LBCO crystal from the same batch after successive irradiations (Fig.~\ref{fig:LBCO_res}). The resistivity curves globally shift to higher values as the irradiation dose is increased, as expected from increased disorder. 
The residual resistivities at zero temperature by extrapolation of the high temperature linear behavior are 2.25, 2.41 and 3.15\,m$\Omega$.cm (which would be 34, 37 and 48\,k$\Omega$ per CuO$_2$ plane, but this includes a significant c-axis contribution, see below).
For LSCO $x=0.10$, adding 2\% Zn causes the in-plane resistivity to more than double\cite{FukuzumiPRLTccupratereducebyZn}, thus the relative change in resistivity for LBCO is comparable to somewhat less than 1\% Zn doping.
At low temperature, the resistivity of our sample drops below the sensitivity of our instruments ($10^{-7}\,\Omega$.cm) below 8\,K for all curves, a few kelvins above the Meissner effect we observed at 4\,K. The value of 8\,K is closer to the 10\,K temperature reported for the c-axis transition to zero resistance\cite{QLiPRL2007,BergPRL2007LBCOTcT3DTKT}.
We attribute this fact and the shape of the resistivity (compared to \cite{QLiPRL2007}) to our sample being slightly misaligned from the in-plane direction (by at most 3.5$^\circ$, from the absolute value of the resistivity) and the extremely large anisotropy of the resistivity $\rho_c/\rho_{ab}$ at low temperature\cite{QLiPRL2007} ($2\times10^3$ to $>10^5$).
But this does not prevent us from observing the superconducting transition, and most importantly the structural and CDW transitions visible in the resistivity of both axes.
In fact, in the derivative of the resistivity we observe a dip at 55.3\,K which matches well with the known orthorhombic to tetragonal structural transition. This transition temperature does not seem to change after the first irradiation, and is smoothed out after the second. A second dip in the derivative, at a temperature immediately below the structural transition, appears to coincide well with the CDW transition. This transition seems to shift from 52.5\,K to 50\,K after the first irradiation, and may reach 35\,K after the second. Finally, the large maximum in resistivity usually associated with the spin stripe order clearly shifts toward lower temperature after irradiation.

\begin{figure}[t!]
\centering
\includegraphics[width=\linewidth]{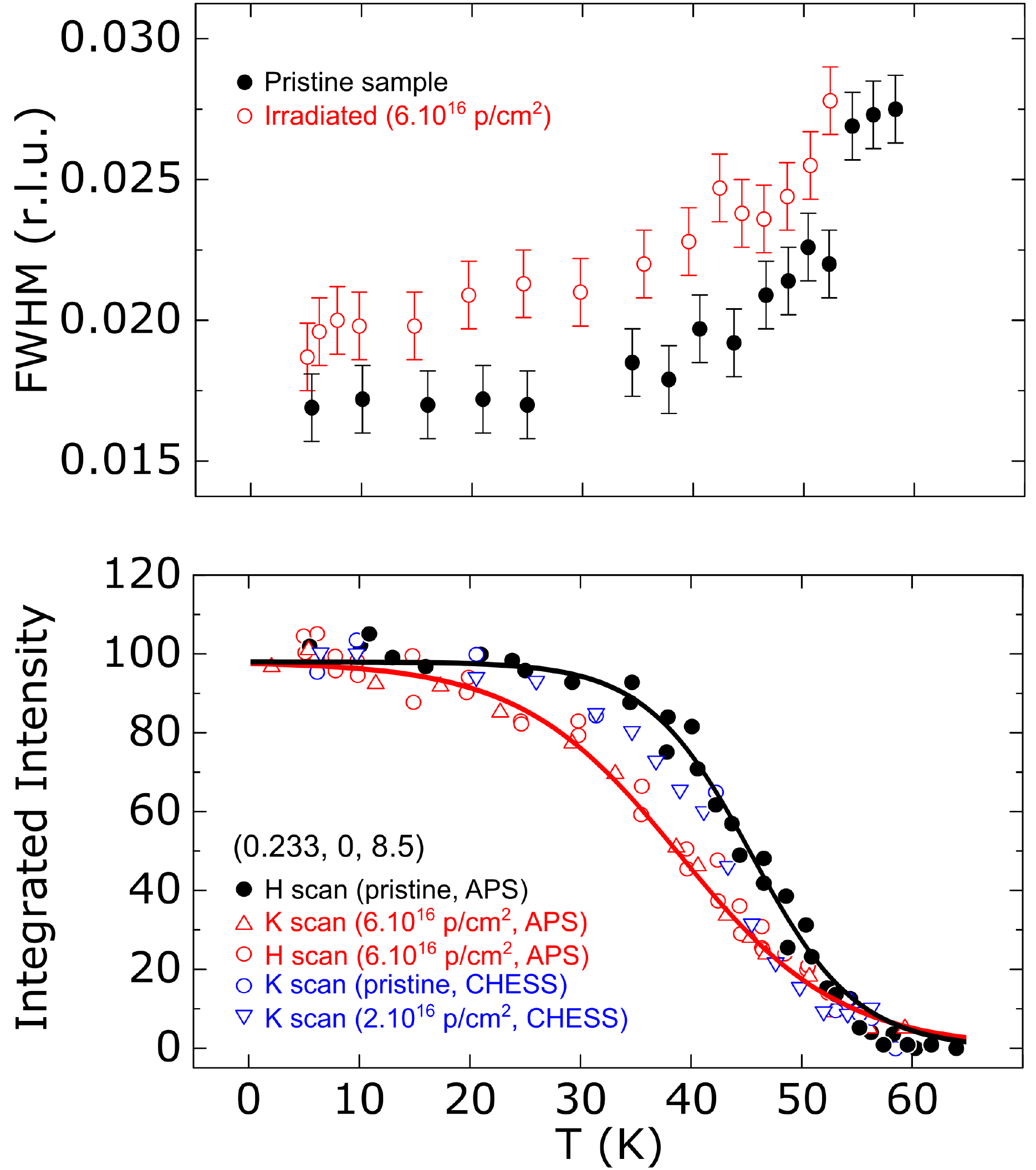}
\caption{\textbf{Irradiation suppresses the CDW in \LBCO}
\textbf{(top)} Temperature dependence of the in-plane full-width at half-maximum (FWHM) of the CDW peak at (0.233,0,8.5) and 8.900 keV, before and after irradiation in the same sample.
\textbf{(bottom)} Temperature dependence of the integrated intensity of the CDW peak at (0.233,0,8.5) for both h-scan and k-scan and for both 8.9 keV (APS) and 27 keV (CHESS). The growth of the order parameter of the CDW state is systematically suppressed by irradiation, but no significant shift of the onset of the CDW is observed within our temperature resolution.
}
\label{fig:LBCO_xray}
\end{figure}
To investigate the effect of irradiation on the CDW we performed x-ray diffraction measurements. Figure~\ref{fig:LBCO_xray} shows the temperature dependence of the integrated intensity of the CDW peak at (0.233,0,8.5) after several irradiations.
In between irradiations we also checked that the sample showed a similar \Tc\ increase (cf. Fig.~\ref{fig:LBCO_curve}).
While the onset of the CDW phase does not appear to change within our temperature resolution, the CDW peak intensity clearly grows more slowly after irradiation. The mid-point of this evolution shifts to lower temperature at a rate of $1.0\,\mathrm{K}/10^{16}\,\mathrm{p/cm^{2}}$.
The precise onset of the CDW order is challenging to determine because of the finite correlation length, which is reduced by irradiation; in addition, dynamic CDW correlations have been reported above the CDW transition temperature\cite{MiaoLBCOCDWnolockin}.
After irradiation, the CDW peak also broadens from 0.017 to $0.02\,a^*$, so that the in-plane long range coherence of the CDW, defined as 1/width, is reduced from 223 to $189\,$\AA.

\section{Discussion}


Irradiation of superconductors typically leads to a decrease of \Tc\ and an increase of the resistivity\cite{AlbenquePRLYBCO7,Albenque,AlloulHirschfeldRMPdefects}. This phenomenology is broadly consistent with expectations based on irradiation-induced pair-breaking scattering\cite{Anderson1959, abrikosov1961aa, Openov98}.  However, the increase of \Tc\ reported here cannot be accounted for in these models.  Recently, it has been shown\cite{ HirschfeldPRL2018} that in a very inhomogeneous spin-fluctuation mediated unconventional superconductor spatial averaging over the inhomogeneous gap can under certain conditions result in an increase of \Tc.  We believe, though, that our samples do not fall into the regime considered in this work. The question then arises whether our results are caused by another extrinsic effect such as an irradiation-induced change in doping level. Irradiation does cause atomic displacements which disrupt the local electronic states and which could, in principle, lead to additional charge transfer and an increase or decrease of \Tc.  
Cooling the samples during irradiation to -10$^\circ$C and the use of beam currents of less than 100 nA\cite{Jia2013protonirrad} prevent beam-induced sample heating and an associated potential loss of oxygen from the sample.  Furthermore, extensive previous work on the effect of irradiation in cuprate superconductors with various particles \cite{ Legris93, Valles89, GOSHCHITSKII1989997} has shown that the Hall coefficient at low temperature does not change upon irradiation even to remarkably high doses.  Thus, to the extent that the Hall number in these materials is a measure of their charge count, the doping level is independent of irradiation.   In addition, our samples afford an intrinsic gauge of the doping level since the incommensurate charge density wave vector is doping dependent\cite{ Hucker2014}.  As shown in SI, the CDW wavevector does not change within the resolution of our measurements: (0.229,0,0.5)$\pm$0.0014 in the pristine sample and (0.231,0,0.5)$\pm$0.0014 in the irradiated sample, respectively, implying that any change in doping is negligible.  Hence, we conclude that our observations are not extrinsic in origin but a reflection of the response of the superconducting order and of the charge order to irradiation induced defects.  Then, in a picture of straight competition between superconducting and charge order an increase of \Tc\ naturally arises if the detrimental effect of irradiation-induced disorder is stronger on the charge order than it is on superconductivity\cite{Mishra2015SDW,Fernandes2012Tcupwithdisorder, Grest82, Psaltakis84}. Indeed, it has been shown that disorder either due to irradiation\cite[p.52]{Mutka1983thesis,Mutka1983PRB_CDW_SC_irrad} or substitution\cite{ Chatterjee2015} strongly suppresses the CDW state either due to pair-breaking \cite{Stiles1976NbSe2} or real-space phase fluctuations \cite{McMillan1975PRB,Mutka1981PRB1TTaS2}.
Hence we conclude that the charge stripe order is strongly competing with bulk superconductivity in \LBCO.

Beyond this competition scenario, the TDO data also provides indications for a possible PDW scenario.
TDO data indeed show enhanced magnetic screening after irradiation below 18\,K when the field is in-plane (lower panel Fig.\ref{fig:LBCO_TDO}), which cannot be explained by the standard effects of disorder.
In the normal state, irradiation should increase resistivity and the skin depth, and in the superconducting state, disorder should also increase $\lambda$.
Therefore, our results points to a reduced $\lambda_{c}$ and enhanced c-axis Josephson coupling between layers as the most likely cause for this enhanced magnetic screening ($\lambda_{ab}$ only changes below 5\,K as shown in Fig.\ref{fig:LBCO_TDO} upper panel).

\begin{figure}
\includegraphics[width=\linewidth]{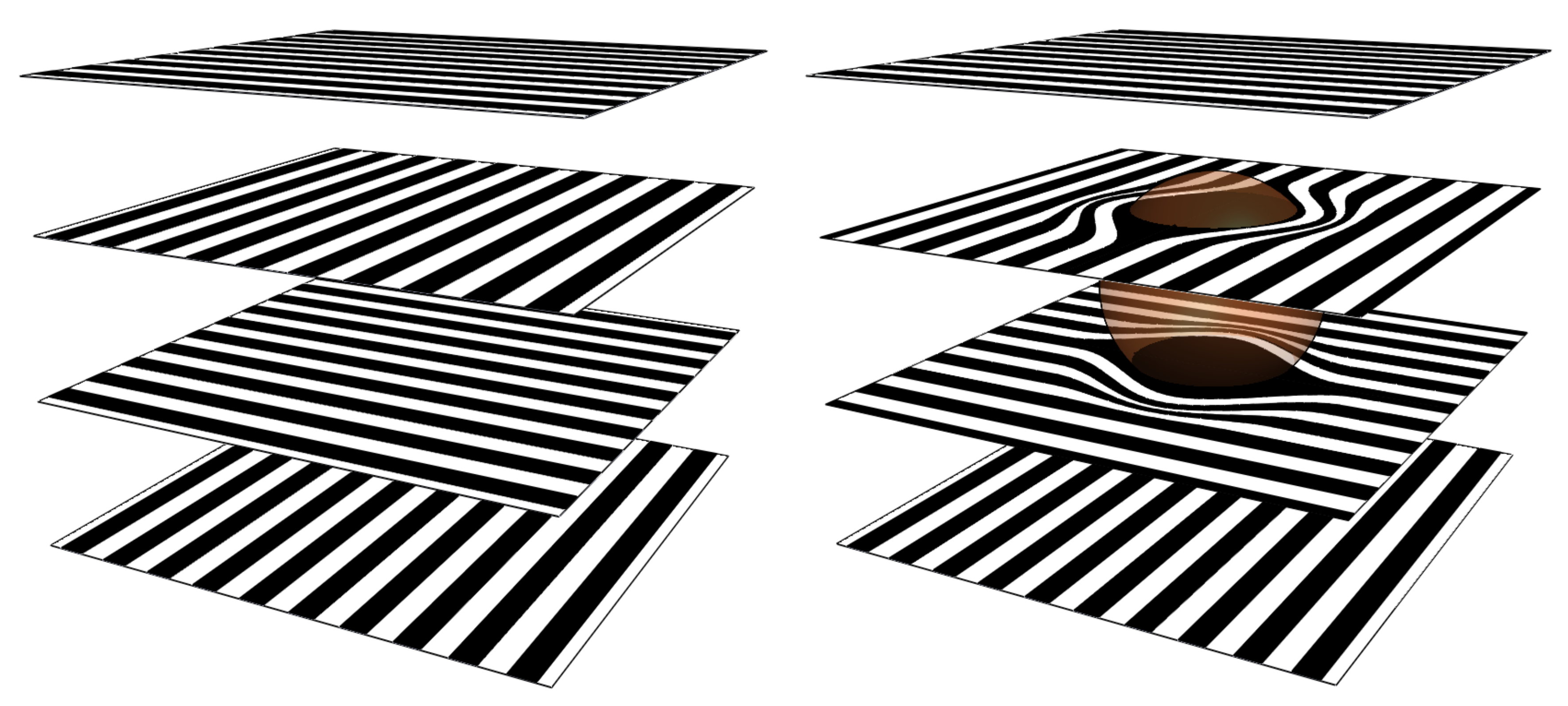}
\caption{\label{fig:disordercoupling} \textbf{Disorder-mediated 3D-coupling re-establishes Josephson coupling}
\textbf{(left)} In an ideal PDW, orthogonal charge-spin stripes in adjacent layers prevent Josephson coupling between layers.
\textbf{(right)} In the presence of disorder, distorted stripes around defects are no more orthogonal which re-establishes Josephson coupling between layers.
}
\end{figure}

It has been proposed that the frustration of the interlayer Josephson tunneling and corresponding suppression of the bulk \Tc\ is the consequence of a pair-density-wave order associated with the stripe order in the CuO$_2$ planes\cite{BergPRL2007LBCOTcT3DTKT} and that this coupling can be recovered upon melting the stripe order with femtosecond laser pulses\cite{RajasekaranCavalleriScience2018,KhannaCavalleri2016}. With perfect PDW order, the spatially alternating sign of the pair wave function leads to a cancellation of the interlayer Josephson coupling.
Therefore, any perturbation of the perfect stripe structure, for example due to irradiation-induced defects as depicted in Fig.~\ref{fig:disordercoupling}, would cause an increase in the Josephson coupling between layers and a rise in \Tc.
Although, the initial evidence for the PDW order in LBCO has been indirect, new experiments are providing direct evidence in support of the occurrence of PDW superconductivity in cuprates\cite{directPDWHarlingen2018,directPDWDavis2018}.
Nevertheless, the PDW scenario and its alternatives are still an actively researched open question.


\begin{figure}[t!]
\includegraphics[width=\linewidth]{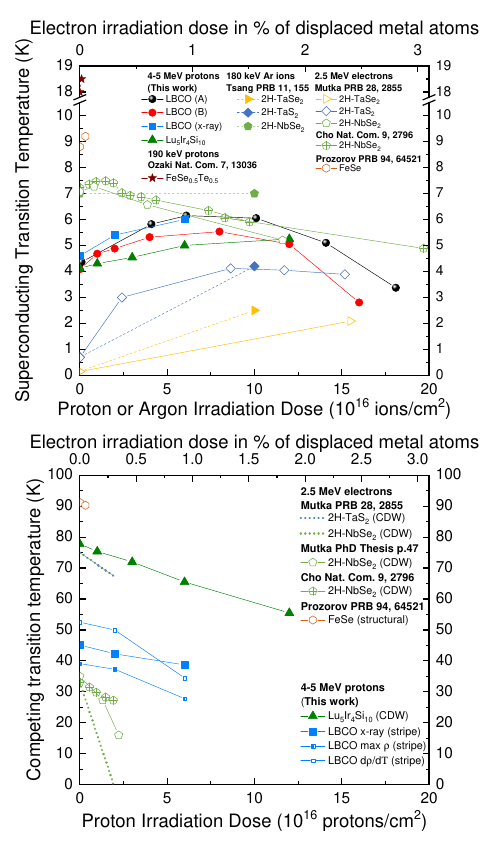}
\caption{\label{fig:LIS_LBCO}  \textbf{Irradiation induced disorder increases \Tc\ and suppresses coexisting orders in several superconductors}
\textbf{(top)}~All known examples of \Tc\ increase after irradiation to the best of our knowledge.
In this work, we report a striking increase of \Tc\ after proton irradiation in the d-wave cuprate superconductor \LBCO,
and a similar, albeit expected, \Tc\ increase in \LIS, a s-wave superconductor with a strong 1D CDW.
Increases of \Tc\ have previously been reported in s-wave dichalcogenides using both argon\cite{TsangArIrradTaSSe2} and electron irradiation\cite{Mutka1983PRB_CDW_SC_irrad,Mutka1983thesis,Cho2018}, and possible hints can be found in iron-based superconductors\cite{OzakiFeSeTeIrrad,ProzorovPRBFeSeIrrad}.
\textbf{(bottom)}~In parallel to the increase of \Tc, a coexisting order was found to be suppressed in some of these compounds. Dotted lines are reported initial trends.
}
\end{figure}

Finally, we point out that competition between superconductivity and CDW or other spatially modulated order is ubiquitous in many families of superconductors.
For instance, in the Fe-based superconductors, a SDW state appears to compete with superconductivity\cite{Fernandes2012Tcupwithdisorder,Mishra2015SDW}, and the underlying magnetism is thought to be a key ingredient for superconductivity. Similarly, in Heavy Fermion compounds such as CeCoIn$_5$, the interaction between SDW and d-wave superconductivity is still actively investigated\cite{Duk2016_PRX_CeCoIn5}. This topic is also a staple of the dichalcogenides superconductors, in which CDW were first discovered. The case for competition is relatively well established in 2H-TaS$_2$ and 2H-TaSe$_2$\cite{Mutka1983PRB_CDW_SC_irrad}. Conversely, there are claims of CDW-enhanced excitonic superconductivity in 1T-TiSe$_2$\cite{KusmartsevaPRL2009TiSe2CDW_SC}. As for 2H-NbSe$_2$, the situation is complex as the two states are only marginally related with potentially both synergy and competition\cite{LerouxPRBNbSe2,Cho2018}. Coexisting density waves and superconductivity are also found in the organic superconductors such as (TMTSF)$_2$PF$_6$\cite{JeromeACSReviewOrganicSC}. To confirm this concept of irradiation as a probe of CDW-superconductivity competition we performed a similar study on crystals of \LIS~(full study to be published elsewhere), which is a s-wave superconductor below 4\,K with a well-characterized quasi-1D CDW below 77\,K\cite{Shelton1986LIS_P_rho_Cp,SinghPRB2005,Leroux2013LIS_SCprop,Mansart2012LIS_femtosec_and_DFT,BettsMigliori2002LIS_RUS}.
The superconductivity and CDW are well known to compete in this compound\cite{Shelton1986LIS_P_rho_Cp,SinghPRB2005}. As can be seen in Fig.\ref{fig:LIS_LBCO}, the increase of \Tc\ and reduction of \TCDW~are strikingly analogous to LBCO.
For completeness, we summarized all irradiation studies on CDW-superconductivity competition in Fig.\ref{fig:LIS_LBCO}. The clarity of the results, but scarcity of data, clearly demonstrate the untapped potential of this technique. Interestingly, beyond its fundamental aspect, a potential application of such increased \Tc\ via irradiation beams could be the direct write of superconducting circuits and Josephson junctions by focused irradiation beams\cite{nanoJJsbyirrad},
which also present the advantage that the superconducting circuits would have much irradiation-induced disorder and thus good vortex pinning properties.

\section{Conclusion}
In summary, we observed a striking \emph{increase} of \Tc\ in \LBCO\ after irradiation, in contrast with the behavior expected of a d-wave superconductor. This increase in \Tc\ is accompanied by a suppression of the CDW, thus evidencing the \emph{strong} competition between CDW and bulk superconductivity in \LBCO. TDO data also points to the restoration of interlayer Josephson coupling frustrated by charge-spin stripes, which would be compatible with a PDW state competing with a uniform d-wave component.


\subsection*{Materials and Methods}
\subsection*{\LBCO Samples}
The samples of LBCO were grown by G.D.Gu. LBCO has a tetragonal crystal structure with space group I4/mmm and lattice parameters a = 3.788 and c = 13.23\,\AA\cite{Abbamonte2005LBCOMottness}.
LBCO samples, cut and polished down to 75 $\mu$m thickness along the c-axis, were repeatedly irradiated with 5 MeV protons using the tandem Van de Graaf accelerator at Western Michigan University. SRIM calculations\cite{Ziegler2010SRIM} predict a penetration depth of 103 $\mu$m at this energy, thus ensuring uniform irradiation damage and negligible protons implantation. We used typical beam currents of 500\,nA spread over one cm$^2$. The irradiation dose was measured by integrating the total current from the irradiation chamber to avoid double counting. The sample was actively cooled to about -10$^\circ$C during irradiation to avoid heat damage. Upstream slits defines a beam spot size of maximum 4.7\,mm diameter at the entrance of the irradiation chamber. A subsequent 1\,$\mu$m-thick gold foil and collimator pair ensures the homogeneity of the incident beam.
\subsection*{SQUID}
The critical temperature of LBCO was measured using the standard zero-field cooled (ZFC) procedure in a custom SQUID (superconducting quantum interference device) magnetometer in a field of 1\,Oersted applied by a copper coil.
\subsection*{Resistivity}
Resistivity measurements were performed in a custom $^4$He cryostat on a 1\,mm\,$\times$\,160\,$\mu$m\,$\times$\,75\,$\mu$m crystal, using a Keithley 2182 voltmeter and a Keithley 6221 current source with a current of 200 $\mu$A.
\subsection*{TDO}
Magnetic penetration depth measurements were performed using a tunnel diode oscillator (TDO) operating at 14-MHz in a $^3$He cryostat.
\subsection*{X-ray}
X-ray diffraction studies were carried on 6-ID-B,C at the Advanced Photon Source. Data were collected in reflection geometry using 8.9 keV photons.
X-ray diffraction studies were carried on A2 at the Cornell High Energy Synchrotron Source (CHESS). Data were collected in transmission geometry using 27 keV photons.
\subsection*{\LIS Samples}
\LIS~is a quasi 1D material with a tetragonal P4/mbm space group symmetry and lattice parameters a = 12.484(1) and c = 4.190(2)\AA\cite{OpagisteLeroux2010LISwhiskers}. The samples of \LIS~were grown at the N\'eel Institute by C.~Opagiste. They grow spontaneously in needle form and have already been characterized elsewhere\cite{OpagisteLeroux2010LISwhiskers,Leroux2013LIS_SCprop}. These needles are high quality single crystals and naturally grow along the c-axis, which is also the axis of the CDW. One sample with as-grown dimensions of $10 \,\mathrm{\mu m} \times 65 \,\mathrm{\mu m} \times 500 \,\mathrm{\mu m}$ $\mathrm{(a\times b \times c)}$ was repeatedly irradiated with 4\,MeV protons at Western Michigan University, using the same irradiation procedure as LBCO. At this energy the projected range of protons is 67\,$\mu$m according to SRIM calculations\cite{Ziegler2010SRIM} ensuring uniform irradiation damage. Resistivity measurements were taken using a Keithley 2182 voltmeter and a Keithley 6221 current source, in a custom $^4$He cryostat.

\section*{Acknowledgements}
The authors declare no conflict of interest.
M.L. and U.W thank Y.-L. Wang for the 3D-coupling schematic.
The experimental study at Argonne National Laboratory was supported by the U.S. Department of Energy, Office of Science, Materials Sciences and Engineering Division.
Theoretical support was provided by V.M. who was supported by the Center for Emergent Superconductivity, an Energy Frontier Research Center, funded by the US Department of Energy, Office of Science, Office of Basic Energy Sciences.
The work at Brookhaven National Laboratory was supported
by the U.S. Department of Energy, Office of Basic Energy Sciences, Division of Materials Sciences and Engineering, under Contract No. DE-SC0012704.
Proton irradiation was performed at Western Michigan University.
This research used resources of the Advanced Photon Source, a U.S. Department of Energy (DOE) Office of Science User Facility operated for the DOE Office of Science by Argonne National Laboratory under Contract No. DE-AC02-06CH11357.
High energy transmission x-ray data was collected on beamline A2 at the Cornell High Energy Synchrotron Source (CHESS) which is supported by the National Science Foundation and the National Institutes of Health/National Institute of General Medical Sciences under NSF award DMR-1332208.

\section*{Authors contributions}
M.L. and V.M. conceived the experiment.
G.G. and J.M.T. grew, prepared and characterized the \LBCO~crystals.
C.O. and P.R grew, prepared and characterized the \LIS~crystals.
M.L., A.K., M.P.S., U.W. and W.K.K. performed the proton irradiations.
H.C., M.L. and U.W. performed the magnetization measurements.
M.P.S. performed the magnetic penetration depth measurements.
M.L. performed the resistivity measurements.
M.L., J.R., Z.I. and U.W. performed the x-ray experiments.
M.L., V.M., P.R., Z.I., U.W., J.M.T. and W.K.K. analysed the data.
M.L., P.R., Z.I and U.W. wrote the manuscript, and all authors contributed to the final version.

\section*{References}
\bibliography{biblio}

\end{document}